# Mitochondrial proteomics: Analysis of a whole mitochondrial extract with two-dimensional electrophoresis

Thierry Rabilloud

CEA/DSV/DRDC/ICH; INSERM U548.  Laboratoire d'Immunochimie, CEA Grenoble, 17 rue des martyrs, F-38054 GRENOBLE CEDEX 9

Email: Thierry.rabilloud@cea.fr

## 1. Introduction

Mitochondria are among the most complex cell organelles, and contain up to 10% of the cell protein content.  Furthermore, they are among the few cell organelles which are separated from the bulk of the cytoplasm by a double membrane (i.e. a double lipid bilayer). As a matter of fact, this is probably linked with the functioning of the mitochondrial energy transducing machinery, which uses a proton gradient across the inner membrane. This proton gradient is built by the oxidative phosphorylation complexes, also named complex I to IV or respiratory complexes, but it also requires a "tight" membrane, in the sense that it must be impermeant even to protons which must re-enter in the mitochondrial matrix only through the ATP synthase (also named complex V) for an efficient ATP production. This implies in turn that this membrane is also impermeant to many other solutes, including those which must be present in the mitochondrial matrix, so that the various biochemical reactions which occur in the mitochondria can take place. This further implies that a whole range of transporters is present in the inner membrane to allow the selective import of these substrates.

Mitochondria are also peculiar in the fact that they possess an autonomous genome. In mammals this genome is almost vestigial and encodes only 13 protein subunits, all very hydrophobic. This implies in turn that a few hundreds of proteins present in the mitochondria are imported, even the mitochondrial ribosomal proteins and the mitochondrial RNA polymerase that are used to produce in situ the mitochondrially-encoded proteins. This import mechanism is quite different from the one used for ER-derived organelles, and has been reviewed elsewhere [1].

Thus, on a protein composition point of view, mitochondria are quite complex and encompass both very soluble proteins (present in the matrix and the intermembrane space) and very hydrophobic membrane proteins, plus membrane proteins of intermediate solubility, such as some subunits of the oxidative phosphorylation complexes or the outer membrane porins. This chemical heterogeneity is a real challenge for the proteomic analysis of mitochondria.

Dysfunction of mitochondria can lead to several disorders of varied severity, ranging from intolerance to an intense effort to perinataly fatal diseases. Progressive mitochondrial dysfunction has also been implicated in the aging process. This has led to interest in comparative mitochondrial proteomics. As many mitochondrial proteins are assembled into complexes of defined stoichiometry (e.g. the respiratory complexes whose structure is sometimes known [2, 3]) it is interesting to reach a fine quantification level which allows to investigate mis-stoichimetries caused by deficient complex assembly. Not all proteomics techniques allow reaching this precision level, and two-dimensional electrophoresis is among the few available choices nowadays. However, this technique is not without drawbacks, especially for hydrophobic proteins [4] and adequate protein solubilization conditions must be used to visualize at least part of the inner membrane-embedded proteins.

The methodological part of this chapter will therefore start with the biological sample (e.g. cultured cells) and detail the mitochondrial preparation, the protein solubilization and the two-dimensional electrophoresis. The mass spectrometry techniques used are quite standard and can be found in any proteomics textbook. A brief outline only will be given in this chapter.

## 2. Materials

### 2.1. Mitochondria preparation

1. Solution A: 10 mM Tris-HCl, pH 6.7, 10 mM KCl, 15 mM $MgCl_2$. This solution is made fresh when needed from stock solutions of Tris buffer (usually 1 M Tris-HCl, pH 7.6), 1 M KCl and 1 M $MgCl_2$. The stock solutions are stable for months at room temperature.

2. Solution B: 2 M sucrose. This solution is prepared with gentle warming (up to 60°C) to help dissolution. Once made, it is kept at 4°C for a few months. Degradation of the solution is indicated by mold or bacterial growth, which is fairly visible on swirling the bottle containing the solution.

3. Solution C: 0.25 M sucrose, 10 mM Tris-HCl, pH 6.7 (at 25°C), $1.5 \times 10^{-4}$ M $MgCl_2$. This solution is prepared on the day of use from solution B and from the stock solutions used for the preparation of solution A.

4. Solution D: 0.25 M sucrose, 10 mM Tris-HCl, pH 7.6, 10 mM EDTA. This solution is prepared on the day of use from solution B, from 1 M Tris buffer and from stock 0.5 M EDTA-NaOH, pH 8.0. The latter solution is prepared by suspending EDTA disodium salt and adding concentrated NaOH (10 M) up to the desired pH, which is close to the dissolution point. Thus care must be taken not to add too much sodium hydroxide. This EDTA stock solution is stable for months at room temperature.

5. Solution E: 0.225 M sucrose, 75mM mannitol, 10 mM Tris-HCl pH 7.6, 1mM EDTA.

*2.2. Mitochondrial proteins solubilization*

1. Solution F: 8.75 M urea, 2.5 M thiourea, 6 mM tris carboxyethyl phosphine and 0.5% (v/v) 3-10 carrier ampholytes (all from Fluka). This solution is difficult to prepare, as water occupies less than 50% of the volume. The most convenient way is to place the capped tube in a bath sonicator and to let sonicate until complete dissolution (occasional tube inverting speeds up the process). Once made, this solution is stored in aliquots at −20°C for up to one year.

*2.3. 2D gel electophoresis*

1. Orange G solution: 2 mg orange G (Fluka) per mL of water. Stable at room temperature for months.

2. Dithiodiethanol (from Fluka): used as supplied, stable for months at room temperature. A slight yellow color may develop and does not prevent its use.

3. Solution G: 6 M urea, 30% (v/v) glycerol, 2.5% (w/v) SDS, 0.125 M Tris-HCl, pH 7.5.

4. Solution H: 130g/L Tris, 0.6 M HCl.

5. Solution I: 150 g/L Tris, 0.6M HCl.

6. Solution J: 30% acrylamide, 0.8% methylene bis acrylamide (to be stored at 4°C). This solution can be purchased ready-made, and this is recommended to avoid handling of monomer powders.

7. Ammonium persulfate solution: 10% (w/v) ammonium persulfate in water, stable for one week at room temperature.

8. Solution K: 2% (w/v) low melting agarose in 0.125 M Tris-HCl, pH 7.5, 0.4% (w/v) SDS, 0.002% (w/v) bromophenol blue.

9. Solution L: 6 g/L Tris, 30 g/L glycine, 1 g/L SDS.

10. Solution M: 3 g/L Tris, 1 g/L SDS, 25 g/L Taurine.

11. Solution N: The silver-ammonia solution is prepared as follows: for ca. 500 mL of staining solution, 475 mL of water are placed in a flask with strong magnetic stirring. First, 7 mL of 1 N sodium hydroxide are added, followed first by 7.5 mL of 5 N ammonium hydroxide (Aldrich) and then by 12 mL of 1 N silver nitrate. A transient brown precipitate forms during silver nitrate addition. It should disappear in a few seconds after the end of silver addition. Persistence of a brown precipitate or color indicates exhaustion of the stock ammonium hydroxide solution. Attempts to correct the problem by adding more ammonium hydroxide solution generally lead to poorer sensitivity.

The ammonia--silver ratio is a critical parameter for good sensitivity [5]. The above proportions give a ratio of 3.1, which is one of the lowest practicable ratios. This ensures highest sensitivity and good reproducibility control of the ammonia concentration through silver hydroxide precipitation. This solution should be prepared at most 30 min before use.

Flasks used for preparation of silver--ammonia complexes and silver--ammonia solutions must not be left to dry out, as explosive silver azide may form. Flasks must be rinsed at once with distilled water, while used silver solutions should be put in a dedicated waste vessel containing either sodium chloride or a reducer (e.g. ascorbic acid) to precipitate silver.

12. Solution O: 2% (w/v) phosphoric acid, 15% (v/v) ethanol and 12% (w/v) ammonium sulfate. Phosphoric acid and ammonium sulfate are added to water (70% of the final volume). Ethanol is added once the salt is dissolved, and the volume is adjusted with water.

*2.4. Protein digestion and analysis by mass spectrometry*

1. 25 mM ammonium bicarbonate.

2. HPLC grade acetonitrile and formic acid.

3. 10 mM dithiothreitol.

4. 55 mM iodoacetamide.

5. Sequencing-grade trypsin.

5. α-cyano-4-hydroxycinnamic acid.

6. Mass spectrometry measurements are carried out on an ULTRAFLEX™ MALDI-TOF/TOF mass spectrometer (Bruker-Daltonik GmbH, Bremen, Germany).

## 3. Methods

### 3.1. Mitochondria preparation

The preparation starts from a cell pellet. As small scale preparations lead to more severe losses and also to lesser mitochondrial purity, it is recommended to start from a billion cells, leading to a few milligrams of mitochondrial proteins. This amount is sufficient to carry out a complete set of comparative mitochondrial proteomics, including several replicate gels and preparative gels for the identification of minor-abundance proteins.

1. After isolation, wash the cells in a standard saline solution (e.g. PBS).

2. Swell the cells in solution A (at least 10 mL of solution A per gram of cell pellet), for 5 min on ice.

3. Break the cells with a motor-driven Potter-Elvejehm homogenizer set at 80-100 rpm. 10 strokes are generally needed to break >80% of the cells, but this may depend on the cell type. This step is carried out in a cold room to limit proteolysis. Protease and phosphatase inhibitors are not used in the lysis buffer because mitochondria are tight organelles, which means that the interior of intact mitochondria is protected from what happens outside. Moreover, many inhibitors do not enter in the mitochondria.

4. Measure the volume of the homogenate, and add $1/7^{th}$ of this volume of cold (4°C) solution B

(*see* **Note 1**).

5. Centrifuge this homogenate at 1,200 g for 5 min at 4°C to get rid of unbroken cells, large debris and nuclei.

6. Collect the mitochondria by centrifugation at 8,000 g for 10 min at 4°C. Resuspend the mitochondrial pellet (by homogenization with 10 strokes of a hand-driven Potter-Elvejem homogenizer) in 20-50 times its volume of solution C.

7. Centrifuge at 1,200 g for 5 min at 4°C, save the supernatant and centrifuge at 8,000 g for 10 min at 4°C.

8. Save the pellet and wash again once by the same procedure, but using solution D (*see* **Note 2**).

9. Store the final mitochondrial pellet in aliquots at –80°C as a concentrated suspension in solution E (estimate the volume of the final pellet and use at most 5 times this volume of solution E to prepare the suspension (*see* **Note 3**). Protein concentration is estimated by a standard protein assay (BCA or Bradford type).

### *3.2. Mitochondrial protein solubilization.*

Protein solubilization for two-dimensional gel electrophoresis is carried out the day of use by mixing at room temperature one volume of mitochondrial suspension, one volume of detergent solution and 8 volumes of solution F (*see* **Notes 4 and 5**). Extraction is carried out at room temperature for 0.5 to 3 h. The solution is then loaded on the isoelectric focusing strip.

### *3.3. Two-dimensional gel electrophoresis*

#### **3.3.1. Isoelectric focusing.**

Because of their simplicity of use and because of their high performance in the analysis of basic proteins (mitochondria are quite rich in basic proteins), the use of immobilized pH gradients

(IPG) strips is strongly recommended. Strips of various pH ranges are commercially available (e.g. from GE Healthcare or from Bio-Rad). Otherwise, immobilized pH gradient plates can be prepared in the laboratory and cut into strips of required width with a paper cutter. This home-made pH gradient preparation is however beyond the scope of this chapter and can be found in adequate textbooks [6]. Nevertheless, commercial or home-made IPG strips are handled the same way. The strips are reswollen in the adequate solution and the protein sample is applied either at this reswelling stage or after reswelling in a sample cup. As a rule of thumb, application by reswelling is preferred, except when it leads to poor resolution, i.e. for basic gradients (e.g. 6-10, 7-11 ranges). However, sample application by reswelling is adequate for wide gradients, even if they extend into the basic pH (e.g. 3-10, 4-12). Both sample application procedures are presented here.

### 3.3.1.1. Sample application by reswelling.

The total amount of solution needed for complete reswelling (i.e. including the sample solution volume) depends on the size of the strip. Commercial strips are 3.3 mm wide, and cast as 0.5 mm thick gels. Thus the strip gel volume in μL is 1.65 x strip gel length (in mm). However, best results are obtained when the gel is reswollen to 1.25-1.3 fold over their initial volume [7]. Thus, the reswelling volume in μL is 2 x strip gel length (in mm).

Practically, prepare a sample dilution solution on the day of use by mixing 8 volumes of solution F, 1 volume of water and 1 volume of 20% detergent solution.

Once the required reswelling volume and the required sample volume are known, dilute the sample up to the reswelling volume with this dilution solution.

To this reswelling sample solution, add (i) 1 μL of Orange G solution and (ii) 0.1 volume of dithiodiethanol (*see* **Note 6**).

Place this complete, colored rehydration solution in the grooved rehydration chamber or in the strip holder, depending on the system used, and let rehydration take place overnight at room temperature, the whole strip plus solution being covered by mineral oil to prevent evaporation. Place then the reswollen gel in the IEF apparatus which is ready for running.

**3.3.1.2. Sample application by cup loading.**

In this case, the ideal gel rehydration volume is the initial one, i.e. (in µL) 1.65 x strip gel length (in mm). The rehydration solution is made by mixing 8 volumes of solution F, 1 volume of dithiodiethanol, 1 volume of 20% detergent solution and 2 µL of Orange G solution per milliliter of rehydration solution. Rehydration takes place overnight at room temperature in the chamber provided by the IEF apparatus supplier. On the day of use, apply the sample at the anodic side of the gel on a plastic cup or in the molded chamber, depending on the apparatus used.

**3.3.1.3. Isoelectric focusing and equilibration**

1. Place the rehydrated strip in the strip holder and apply the sample anodically if necessary. This is required when alkaline pH gradients (e.g. 6-12, 7-11) are used.

2. Cover the strip and the sample with mineral oil and connect the power supply.

3. It is advisable to use a thermostated IEF apparatus to guarantee the constancy of the spot position in the 2D pattern [8]. The strips can be run at any temperature above 10°C to avoid urea crystallization. The strips are usually run at 22°C to avoid any precipitation of urea-detergent complex.

4. To avoid any overheating, even local ones, it is recommended to use a voltage-controlled migration program. For a wide pH gradient (3 pH units or more) which is 16 to 20 cm long, the following program is used: 100 V for 1 h, 300 V for 3 h, 1000 V for 1 h, and 3500 V for 18 h or more. To adjust the migration for each condition, the following rule of thumb can be applied: most proteins have reached their steady-state position after 100 V x h/cm$^2$, where the cm$^2$ means the square of the strip length in cm. For example, a 20–cm-long strip needs at least 20x20x100 Vh i.e. 40,000 Vh. However, as most gradients are stable over time, more Vh can be applied without any problem.

5. After the IEF migration has been completed, remove the mineral oil, and equilibrate the strips for 20 min in solution G. 5-10 mL of solution G are used per strip. The equilibrated strips can then be used immediately for the second dimension or frozen in the equilibration solution at –20°C. Frozen strips are stable for a few weeks at –20°C. They are thawed in a water bath at 20-30°C and used when necessary.

### 3.3.2. SDS electrophoresis

#### 3.3.2.1. Gel casting

In addition to being rich in basic proteins, mitochondria are rich in low-molecular-weight proteins, and many subunits of the respiratory complexes are below 15 kDa. This makes the standard Laemmli system not optimal for the resolution of mitochondrial proteins, and it is recommended to use the recently-introduced Tris-Taurine system [9] (*see* **Note 7**).

1. 1.5-mm-thick gels are routinely used. For a 160x200x1.5 mm gel, 60 mL of gel mix are prepared. This gel mix is optimized for the molecular weight range that needs to be investigated.

2. For a 20-200 kDa range, the gel mix is made of 10 mL of solution H, 20 mL of solution J and 30 mL of water. For a 5-200 kDa range, which provides resolution of the low-molecular-weight proteins at the expense of the compression of proteins above 35 kDa, the gel mix is composed of 10 mL of solution I, 22.5 mL of solution J and 27.5 mL of water.

3. Initiate polymerization by the sequential addition of 20 μL of TEMED (tetramethyl ethylene diamine) and 400 μL of ammonium persulfate solution (*see* **Note 8**).

4. Cast the gels between the plates (5 mm free of gel mix are left at the top of the gel cassette) and overlay with 0.8 mL of water-saturated 2-butanol. Polymerization should occur within 30 min.

5. It is recommended to cast the second dimension gels the day before their use for a complete and uniform polymerization. Once polymerized, remove the gels from the casting chamber, remove the butanol and replace with water, and store the gels assemblies in a closed polyethylene box. To avoid glass plate sticking, separate each gel assembly from its neighbors by a plastic sheet (polycarbonate plastic sheets from Bio-Rad).

### 3.3.2.2. Strip transfer and gel running

1. Place the second dimension gel assembly on its stand.

2. Catch the equilibrated strip with tweezers at one end. The use of inverted tweezers which hold the strip without hand pressure is quite convenient.

3. Clip the excess plastic and gel at the free end (the one not covered by the tweezers) with scissors.

4. Pour 0.8 mL of molten agarose (solution K) on the top of the second dimension gel, and put the strip in place (clipping of the excess plastic and gel at the site of the tweezers releases the strip in place). Care must be taken to eliminate any bubble between the top of the second dimension gel and the strip.

5. Allow 10 min for the agarose to set (*see* **Note 9**) and secure the gel in the gel tank.

6. Fill the lower chamber of the tank with buffer L and the upper chamber with buffer M (*see* **Note 10**).

7. Run the second dimension gels at 10°C (thermostated) for 1 h at 25 V, then at 12 W/gel until the bromophenol blue front reaches the bottom of the gel.

**3.3.3. Spot visualization**

Two main types of spot visualization are used in such proteomics experiments. Silver staining is used in the initial phases of the study, for example to set the conditions and to perform comparative experiments. The rationale for using silver staining is based on its sensitivity, as a gel showing more than 1000 protein spots can be obtained with 0.1 mg of total mitochondrial proteins. Such "analytical" gels can be used for image analysis but also for spot excision and subsequent protein identification with mass spectrometry. However, silver-stained gels can be deceptive in spot identification, because (i) small and weak silver-stained spots contain very small amounts of proteins and (ii) the silver staining process results in peptide losses in the mass spectrometry identification process [10].

Two main processes can be used for silver staining. The silver nitrate process works well for acidic proteins, but less well for basic proteins. In addition, peptide losses are often important. The silver-ammonia process works nicely for basic proteins, but frequently gives artifacts (weak or hollow or negative spots) in the acidic range. However, the peptide losses are generally lesser with this process. For optimal performance, silver-ammonia staining requires home-made gels cast with sodium thiosulfate (*see* **Note 10**)

Both processes are given below.

**3.3.3.1. Fast Silver Nitrate Staining (*see* Note 11 first)**

1. Fix the gels (1 h + overnight) in acetic acid/ethanol/water, 5/30/65 (v/v/v).

2. Rinse in water for 4 x 10 min.

3. To sensitize, soak gels for 1 min (1 gel at a time) in 0.8 mM sodium thiosulfate.

4. Rinse 2 x 1 min in water.

5. Impregnate for 30-60 min in 12 mM silver nitrate (0.2 g/L). The gels may become yellowish at this stage.

6. Rinse in water for 5-15 s.

7. Develop image (10-20 min) in 3% (w/v) potassium carbonate containing 250 µL of formalin and 125 µL of 10% (w/v) sodium thiosulfate per liter.

8. Stop development (30-60 min) in a solution containing 40 g of Tris and 20 mL of acetic acid per liter.

9. Rinse with water (several changes) prior to drying or densitometry.

### 3.3.3.2 Ammoniacal Silver staining (*see* Note 11 first).

Thiosulfate is added to the gel during polymerization (*see* **Note 10**). After electrophoresis, proceed as follows for silver staining:

1. Fix in acetic acid/ethanol/water, 5/30/65 (v/v/v) containing 0.05% (w/v) 2-7 naphtalene disulfonate (Acros) for 1 h.

2. Fix overnight in the same solution.

3. Rinse 6 x 10 min in water.

4. Impregnate for 30-60 min in the ammoniacal silver solution (solution N).

5. Rinse 3x 5 min in water.

6. Develop image (5-10 min) in 350 µM citric acid containing 1 mL formalin per liter.

7. Stop development in acetic acid/ethanolamine/water, 2/0.5/97.5 (v/v/v).. Leave in this solution for 30-60 min.

8. Rinse with water (several changes) prior to drying or densitometry.

### 3.3.3.3. Protein spots de-staining

When protein identification is planned on spots stained with silver, a destaining step is required for better results. To maximize peptide recovery by in-gel proteolytic digestion, this destaining step should be performed the same day as silver staining [10].

Destaining proceeds as follows [11]:

1. Prepare a stock solution of potassium ferricyanide (30 mM in water) and a solution of sodium thiosulfate (100 mM in water). Just before use, mix equal volumes of the two solutions and cover the spots with the resulting mix.

2. Destain the spots for 5 min at room temperature.

3. Remove the destaining solution, rinse 3x5 min with water.

4. Soak the spots for 20 min in ammonium bicarbonate (200 mM).

5. Remove the bicarbonate solution and rinse 3x5 min in water.

For the less abundant protein spots whose identification fails from silver-stained gels, or when maximal sequence coverage is desired (for example for assignment of modification sites) more heavily loaded gels are needed (0.5 to 1 mg protein loaded on the strip). These gels are usually stained with colloidal Coomassie blue, which is far less sensitive than silver staining but gives much better sequence coverage.

### 3.3.3.4. Colloidal Coomassie blue staining

1. After electrophoresis, fix the gels 3 x 30 min in ethanol/water, 30/70 (v/v) containing 1.7% (w/v) phosphoric acid. This fixation can also proceed with a 1-hour bath followed by an overnight bath.

2. Rinse 3x 20 min in 1.7% (w/v) phosphoric acid.

3. Equilibrate for 30 min in solution O.

4. Without removing the solution surrounding the gels, add 1% (v/v) of a solution containing 20 g of Brilliant Blue G per liter (dissolved in hot water, stable for months at room temperature). Let the staining proceed for 24 to 72 h.

5. If needed, destain the background with water. Avoid alcohol-containing solutions.

*3.4. Mass spectrometry*

The details of mass spectrometry analysis are not fully in the scope of this chapter, as rather classical procedures are used. The detailed procedures used are mentioned for information.

**3.4.1. In-gel digestion**

1. Wash the spots (a robotic device can be used) with 0.1 mL of 25 mM ammonium bicarbonate for 8 min.

2. Remove the bicarbonate and shrink the gel pieces 3x8 min with pure acetonitrile.

3. Remove the acetonitrile and dry completely the gel pieces in a vacuum centrifugal concentrator (e.g. a SpeedVac).

4. Cover the gel pieces with 0.1 mL of 10 mM dithiothreitol in 25 mM ammonium bicarbonate, and break disulfide bridges by incubating at 50°C for 1 hour.

5. Add 0.1 mL of 55 mM iodoacetamide in 25 mM ammonium bicarbonate, and alkylate the thiol groups at room temperature for 1 day (in the dark).

6. Wash alternatively with 25 mM ammonium bicarbonate and acetonitrile (5 min each). Repeat this double washing three times.

7. Dry completely the gel pieces in a vacuum centrifugal concentrator (e.g. a SpeedVac).

8. Estimate the dried gel volume and add three times this volume of trypsin solution (12.5 mg/L in 25 mM ammonium bicarbonate). Let digestion proceed overnight at 35°C.

9. Dry partially the gel pieces for 5 min in a vacuum centrifugal concentrator (e.g. a SpeedVac).

10. Add 5 μL of water/acetonitrile/formic acid, 35/60/5 (v/v/v) and extract the peptides with sonication (bath sonicator) for 5 min. Recover the liquid phase and repeat this extraction once.

### 3.4.2. Mass spectrometry measurements

The MALDI-TOF/TOF instrument is used at a maximum accelerating potential of 20 kV and operated in reflector positive mode. Sample preparation is performed with the dried droplet method using a mixture of 0.5 μL of sample with 0.5 μL of matrix solution. The matrix solution is prepared from a saturated solution of α-cyano-4-hydroxycinnamic acid in $H_2O$/acetonitrile, 50/50 (v/v/v), diluted 3 times in $H_2O$/acetonitrile, 50/50 (v/v). Internal calibration is performed with tryptic peptides resulting from autodigestion of trypsin (monoisotopic masses at m/z = 842.51 Da, m/z = 1045.56 Da, m/z = 2211.11 Da).

Monoisotopic peptide masses are assigned and used for database searches using the search engines MASCOT (Matrix Science, London, UK) and Aldente (www.expasy.org). All proteins present in Swiss-Prot are used without any pI and Mr restrictions. The error on peptide mass measurement is limited to 50 ppm, one possible cleavage site missed by trypsin is accepted.

## 3.5. A few results

In order to illustrate the methods presented in this chapter, some results obtained on mitochondria prepared from human cultured cells are shown. Figure 1 shows the resolving power of two-dimensional electrophoresis of mitochondria prepared from HeLa cells. The resolution in the low molecular weight region, as well as in the basic range, should be noted. These methodological improvements have allowed a better coverage of the mitochondrial proteome [12, 13]

Figure 2 shows a typical comparative experiment, between mitochondria extracted from normal cells (gel A) and mitochondria extracted from cells devoid of mitochondrial DNA (gel B). Some of the differentially expressed spots have been excised from the silver-stained gels, digested and analyzed by mass spectrometry to obtain their peptide mass fingerprint. The results in terms of protein identifications are shown in table 1. More complete results can be found in the literature [14].

## 4. Notes

1. The addition of solution B is intended to restore the osmolarity of the solution to a level close to the one present in cells. This limits mitochondria and nuclei breakage in the following steps, thereby increasing the yield and purity of mitochondria. However, solution B is dense and viscous, and thorough mixing is needed. This is usually achieved by inverting the capped tube and/or by gentle vortex mixing.

2. The washing steps are critical for increasing the purity of the mitochondria. Pure mitochondria are best obtained by density gradient centrifugation or free-flow electrophoresis [15], but these procedures have low yields and are best suited for large samples (e.g. bovine tissues or yeast cultures). The low/high speed procedure presented here is a good compromise between yield and purity, and is adequate for mitochondria isolated from cultured cells. The first wash in solution C removes the last contaminating nuclei, and the second wash in solution D eliminates most of the contaminating ribosomes. The first wash in solution C must not be skipped, as any nuclei remaining in the pellet suspended in solution D will burst because of the presence of EDTA. Exploded nuclei may lead to considerable increase in viscosity and inefficient washing, leading to heavily contaminated mitochondria.

3. Mitochondrial suspensions are more stable when stored concentrated. A good procedure is to resuspend the final pellet in twice its estimated volume of solution E. Suspending is achieved by vortex mixing and pipetting. The bulk of the suspension is saved in a tube, and an equal volume of solution E is added to the tube containing the remnants of the pellet. Thorough suspending is carried out again, and this new suspension is combined to the previous one. This process is carried out in a cold room.

4. There is no single detergent allowing the optimal extraction and thus visualization of all classes of proteins under the conditions prevailing in isoelectric focusing. Soluble proteins are best analyzed with CHAPS, while membrane proteins are poorly if at all solubilized by this detergent. Analysis of membrane proteins requires other detergents such as Brij 56 or dodecyl maltoside which are not equivalent in their solubilization patterns [16]. The choice of detergent will then depend on the focus of the study or on the amount of sample, which will allow or not series of experiments to be carried out with different detergents.

5. In some cases, this procedure will lead to too dilute a protein solution (e.g. for heavily loaded preparative gels or for cup loading). If this is the case, collect first the mitochondria in an Eppendorf-type tube by centrifugation (10,000g for 10 min at 4°C), and then suspend the pellet in an equal volume of 10% detergent solution plus four volumes of solution F.

6. The orange G is used as a tracker dye to check for any lack of electrical contact which would prevent protein migration at the isoelectric focusing stage. The dye must migrate to the anode and collect in a small zone close to the anode. Dye remaining over a large portion of the strip indicates a migration problem (electrical contact problem or too high salt concentration).

Dithiodiethanol (used as supplied) has been shown to increase resolution in the basic portion of the IPG gels [17] and simplifies equilibration between the IPG and SDS dimension. It can be used with any pH range.

7. The Tris-Taurine system uses both the pH and the acrylamide concentration to control the speed of the moving boundary (visualized as the bromophenol blue front). This speed controls in turn the speed of the proteins that can comigrate with this dye front, and thus the lower limit (in molecular weight) of the proteins that can be resolved in the gel. This strong dependency of the boundary speed upon pH explains the way of preparing the gel buffer. These recipes have been empirically optimized, and their way of preparation is designed to give the best reproducibility over time.

8. For optimal results, alterations must be made at the level of gel casting when silver ammonia staining is to be performed. Thiosulfate is added at the gel polymerization step. Practically, the initiating system is composed of 1 μL of TEMED, 7 μL of 10% (w/v) sodium thiosulfate solution and 8 μL of 10% (w/v) ammonium persulfate solution per mL of gel mix. This ensures correct gel formation and gives minimal background upon staining.

9: Low melting agarose is used as it will leave more time to put the strip in place before the agarose gel sets. This can be a problem in summer in warm labs, where the temperature is close to the setting temperature of low melting agarose. If this is the case, the most convenient way is to place the gel assemblies in the fridge for 1 h prior to strip transfer. This will secure agarose gel setting.

10. Two different buffers can be used in gel tanks where the cathode and anode chambers do not communicate on a fluidic point of view (e.g. Bio-Rad Protean chambers). In this case, the lower chamber is usually much bigger than the upper one, and a cheaper Tris glycine buffer can be used in the lower chamber. Taurine is needed only in the cathode (upper) chamber for the system to operate. In "submarine" type gel tanks (e.g. Bio-Rad dodeca cell) the Taurine buffer (solution M) must be the only electrode buffer used.

11. General practice for silver staining. Batches of gels (up to five gels per box) can be stained. For a batch of three to five medium-sized gels (e.g. 160 x 200 x 1.5 mm), 1 L of the required solution is used, which corresponds to a solution/gel volume ratio of 5 or more; 500 mL of solution is used for one or two gels. Batch processing can be used for every step longer than 5 min, except for image development, where one gel per box is required. For steps shorter than 5 min, the gels should be dipped individually in the corresponding solution.

For changing solutions, the best way is to use a plastic sheet. The sheet is pressed on the pile of gels with the aid of a gloved hand. Inclining the entire setup allows the emptying of the box while keeping the gels in it. The next solution is poured with the plastic sheet in place, which prevents the solution flow from breaking the gels. The plastic sheet is removed after the solution change and kept in a separate box filled with water until the next solution change. This water is changed after each complete round of silver staining. The above statements are not true when gels supported by a plastic film are stained. In this case, only one gel per dish is required. A setup for multiple staining of supported gels has been described elsewhere [18].

When gels must be handled individually, they are manipulated with gloved hands. The use of powder-free, nitrile gloves is strongly recommended, as powdered latex gloves are often the cause of pressure marks. Except for development or short steps, where occasional hand agitation of the staining vessel is convenient, constant agitation is required for all the steps. A reciprocal ("ping-pong") shaker is used at 30-40 strokes per min.

Dishes used for silver staining can be made of glass or plastic. It is very important to avoid scratches in the inner surface of the dishes, as scratches promote silver reduction and thus artifacts. Cleaning is best achieved by wiping with a tissue soaked with ethanol. If this is not sufficient, use instantly prepared Farmer's reducer (50 mM ammonia, 0.3% potassium ferricyanide, 0.6% sodim thiosulfate). Let the yellow-green solution dissolve any trace of silver, discard, rinse thoroughly with water (until the yellow color is no longer visible), then rinse with 95% ethanol and wipe.

Formalin stands for 37% formaldehyde. It is stable for months at room temperature. However, solutions containing a thick layer of polymerized formaldehyde must not be used. Never put formalin in the fridge, as this promotes polymerization. 95% ethanol can be use instead of absolute ethanol. Do not use denatured alcohol. It is possible to purchase 1 M silver nitrate ready-made. The solution is cheaper than solid silver nitrate on a silver weight basis. It is stable for months in the fridge.

Last, but not least, the quality of water is critical. Best results are obtained with water treated with ion exchange resins (resistivity higher than 15 MΩ/cm). Distilled water gives more erratic results.

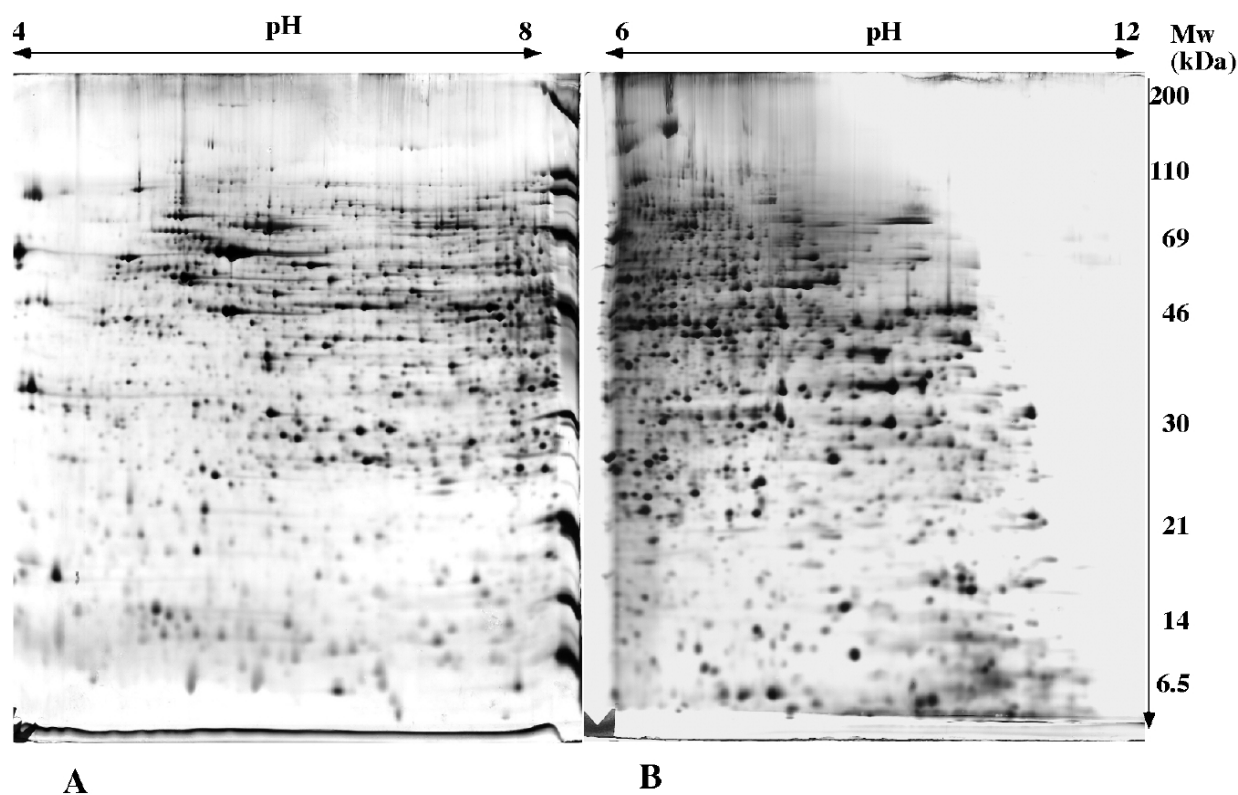

**Figure 1**: two-dimensional separation of mitochondrial proteins.

Mitochondrial proteins obtained from HeLa cells are separated on the gels. 0.12 mg of total mitochondrial protein has been loaded on each gel. The second dimension gel is a 11.5% gel cast with the alkaline gel buffer (solution I).

A: pH 4-8 linear pH gradient in the first dimension. Sample application by in-gel rehydration. Spot visualization with silver nitrate staining.

B: pH 6-12 linear pH gradient in the first dimension. Sample application by cup loading. Spot visualization with silver-ammonia staining.

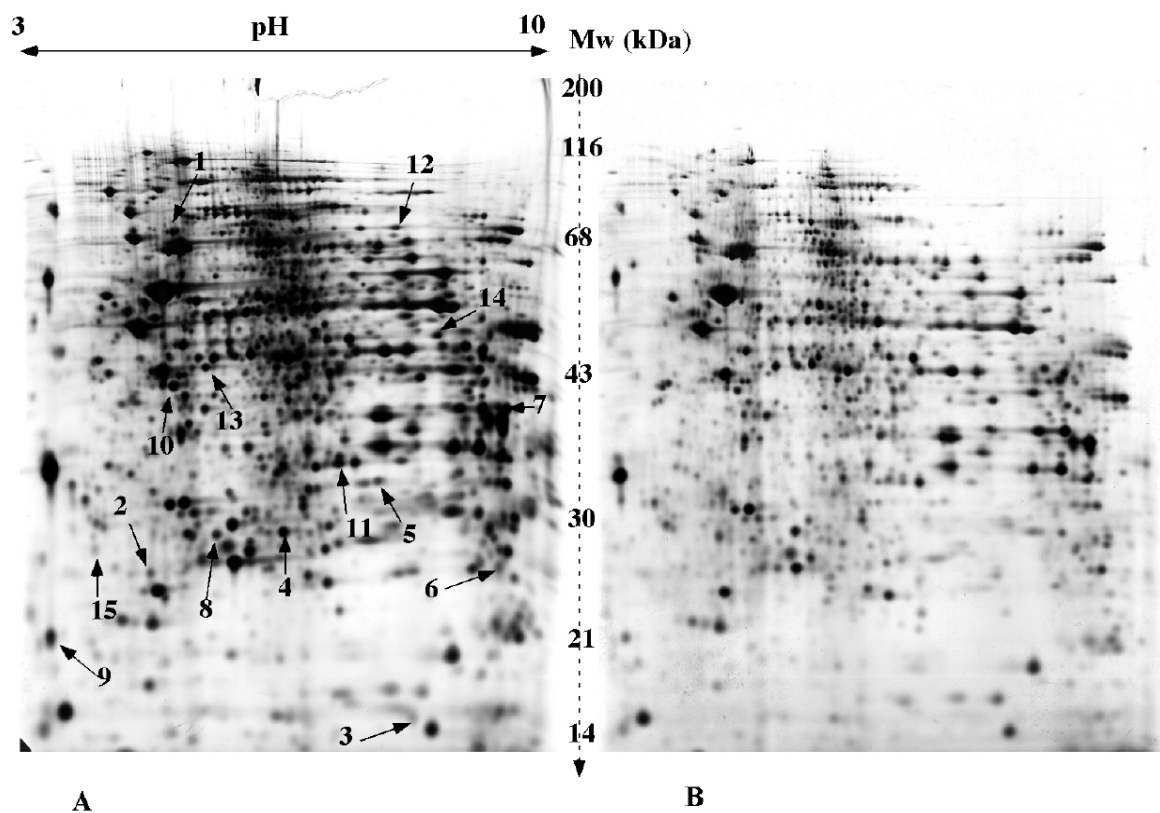

**Figure 2**: comparative proteomics on mitochondrial proteins

Mitochondrial proteins obtained from 143B cells or Rho0 cells are separated on the gels. 0.12 mg of total mitochondrial protein has been loaded on each gel. pH gradient: linear 3-10. Sample application by in-gel rehydration. The second dimension gel is a 10% gel cast with the standard gel buffer (solution H). Spot visualization by silver nitrate staining.

A: mitochondrial proteins extracted from 143B cells (normal mitochondrial DNA)

B: mitochondrial proteins extracted from 143B Rho0 cells (cells devoid of mitochondrial DNA)

Some differentially-expressed spots have been excised, destained, and submitted to protein identification by mass spectrometry (peptide mass fingerprinting). The results are shown in table 1.

**Table 1**: Protein identification results

The protein identification results obtained from the spots excised from the gels of figure 2 are shown. The quantitative variations between 143B and Rho0 cells have been obtained by quantitative image analysis, using the Melanie software and the normalization of protein intensities in ppm of the total.

| number on gels | Acc number | name | pI | Mw (kDa) |
|---|---|---|---|---|
| 1 | P28331 | NADH-ubiquinone oxidoreductase 75 kDa subunit | 5.42 | 77 |
| 2 | O00217 | NADH-ubiquinone oxidoreductase 23 kDa subunit | 5.10 | 20.3 |
| 3 | Q9P0J0 | NADH-ubiquinone oxidoreductase B16.6 subunit | 8.24 | 16.5 |
| 4 | P47985 | Ubiquinol-cytochrome c reductase iron-sulfur subunit | 6.3 | 21.6 |
| 5 | Q9Y399 | Mitochondrial 28S ribosomal protein S2 | 7.35 | 28.3 |
| 6 | Q9Y3D9 | Mitochondrial ribosomal protein S23 | 8.94 | 21.7 |
| 7 | Q9UGM6 | Tryptophanyl-tRNA synthetase | 8.99 / 8.29 | 37.9 / 35.1 |
| 8 | O75431 | Metaxin 2 | 5.9 | 29.7 |
| 9 | Q9NS69 | Mitochondrial import receptor subunit TOM22 | 4.27 | 15.4 |
| 10 | Q9UJZ1 | stomatin-like protein 2 | 5.26 | 35.0 |
| 11 | P13804 | Electron transfer flavoprotein alpha-subunit | 7.1 | 32.9 |
| 12 | P49748 | Acyl-CoA dehydrogenase, very-long-chain specific | 7.74 | 66.1 |
| 13 | Q9P2R7 | Succinyl-CoA ligase [ADP-forming] beta-chain | 5.64 | 43.6 |
| 14 | Q6YN16 | Hydroxysteroid dehydrogenase | 6.31 | 42.5 |
| 15 | Q96EH3 | Chromosome 7 open reading frame 30 | 4.87 | 21.6 |